\documentclass{article}

\usepackage{arxiv}

\usepackage[utf8]{inputenc} 
\usepackage[T1]{fontenc}    
\usepackage{hyperref}       
\usepackage{url}            
\usepackage{booktabs}       
\usepackage{amsfonts}       
\usepackage{nicefrac}       
\usepackage{microtype}      
\usepackage{lipsum}
\usepackage{graphicx}
\graphicspath{ {./images/} }
\title{Resource efficient single photon source based  on active frequency multiplexing}
\author{
 Serge Massar$^1$ and St\'ephane Clemmen$^{1,2,3,4}$\\
  $^1 $Laboratoire d’Information Quantique,   Universit\'e Libre de Bruxelles (ULB) CP 224,   1050 Brussels,   Belgium \\
  $^2 $ OPERA-Photonique CP 194/5, Universit\'e Libre de Bruxelles (ULB), 1050 Brussels, Belgium \\
  $^3$ Photonics Research Group, Department of Information Technology, \\Ghent University-IMEC, Technologiepark 126, 9052 Ghent, Belgium \\
   $^4$ Center for Nano- and Biophotonics, Ghent University, 9052 Ghent, Belgium \\
 $*$ \textit{sclemmen@ulb.ac.be} 
}
\begin{document}
\date{\vspace{-5ex}}
\maketitle

\begin{abstract}
We propose a new  single photon source based on the principle of active multiplexing of heralded single photons which, unlike previously reported architecture, requires a limited amount of physical resources. We discuss both its feasibility and  the purity and indistinguishability of single photons as function of the key parameters of a possible implementation.
\end{abstract}


Single photon sources are an essential building block for optical quantum processing and communications. Features that are associated to the high quality of a single photon source are
\cite{lounis_single-photon_2005}: their purity, i.e. the ability for the source to emit on demand one and only one photon;  their indistinguishably, i.e. how identical to each other are photons consecutively emitted;  the possibility to make many identical sources; and the practicality (compactness, cost, aging, operating temperature, etc). Many other features  are important but do not make for an absolute good or bad source such as the duration/linewidth of the photon wavepacket, its temporal/spectral shape, its polarization, its spatial profile,  the confinement in a single mode waveguide. 

\begin{figure}
\begin{center}
\includegraphics[width=11cm]{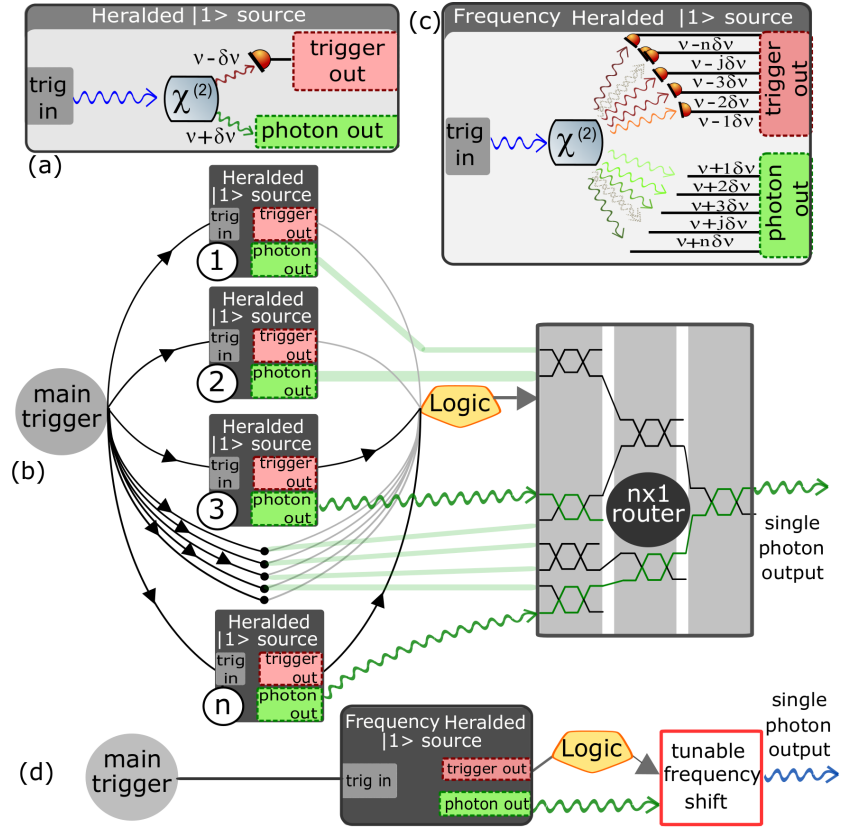}
\end{center}
\caption{(a) A heralded source of single photon. (b) Principle of active multiplexing of heralded single photons (implemented in the spatial domain). In the presented figure, two photons are heralded (black arrows) and distributed (green arrows) while other channels are empty (light lines). The logic reconfigures the router to exit only one heralded photon. (c) Principle of a \textit{frequency} heralded source of single photon: the trigger out indicates not only the detection of a photon but also its wavelength.  (d)  Principle of active multiplexing of frequency heralded single photons. The frequency shifter is implemented using a frequency tunable  wave mixing process.
}
\label{fig:1}
\end{figure}

While single photon sources based on single emitters are being improved regularly
, these sources remain intrinsically limited in term of tunability.  Heralded photons~\cite{fasel2004high} (see fig. \ref{fig:1}a) originating from parametric sources of photon pairs are often used in place of such deterministic sources because they typically offer good indistinguishably and spectral-temporal properties that can be tailored to a large extend.  
Their intrinsically limited purity (less than 25\%~\cite{christ_limits_2012}) can be enhanced via an architecture known as active multiplexing~\cite{migdall_tailoring_2002,castelletto2008heralded,kaneda_time-multiplexed_2015,bonneau2015effect,xiong_active_2016,mendoza_active_2016,grimau_puigibert_heralded_2017}.
A spatial implementation of that principle is illustrated  in fig. \ref{fig:1}(b). It relies on a large number $N$  of heralded sources. Individually, each heralded source is characterised by the trigger probability $p_{\textrm{trig}}^{\textrm{MUX}}$ and the probability $p_{\textrm{single}}$ for the heralded state to be a single photon. The probability that at least 
one among the $N$ sources  emits a single photon is 
\begin{equation}
p_1 = p_{single}  \left( 1- (1 - p_{\textrm{trig}}^{\textrm{MUX}})^N\right)
\label{Eq:p1}
\end{equation} 
So, even for moderate values of  $p_{\textrm{trig}}^{\textrm{MUX}}$, a large enough $N$ brings asymptotically to $1$ the second factor. However this can be resource demanding, with values of $N$  in the hundreds required.
 In addition, the losses usually scales with $N$. 
For instance in a spatial implementation using a tree of $N \times 1$ routing element the losses are proportional to $\log_2(N)$. 
Even if this scaling can be slightly improved using clever logic~\cite{gimeno2017relative}, the losses of the active routing element remain an important limitation.

It has been recently reported~\cite{joshi_frequency_2018} that the unfavourable scaling  with $N$ could be avoided in a frequency multiplexing implementation by using a frequency heralded  photon source as figure \ref{fig:1}(c). This replaces the multitude of distinct sources in (b) by a multitude of frequency bins within a unique source. Figure \ref{fig:1}(d) shows how this source can be combined with a frequency shifter to perform the same task as the  $N \times 1$ routing element in the spatial implementation.  In  \cite{joshi_frequency_2018} the number of single photon detectors is  equal to $N$ as in the case of the spatial implementation illustrated in figure \ref{fig:1}(c) and the frequency shifter is  made of $N$ lasers at $N$ different wavelengths, one of which is fired upon the detection of the heralding single photon. This first demonstration was implemented with $N = 3$. This small number was chosen because the architecture is demanding in terms of physical resources, but was  too small to reach a high probability of success  $p_1$. While the demonstration was a  success, the scaling of the architecture to $N$ on the order of tens or hundreds seemed challenging. Further improvements were proposed \cite{PatentGrice2019,,hiemstra2020pure}, but still facing some difficulties scaling up to a large $N$.   shifter. 

\begin{figure}
\begin{center}
\includegraphics[width=11cm]{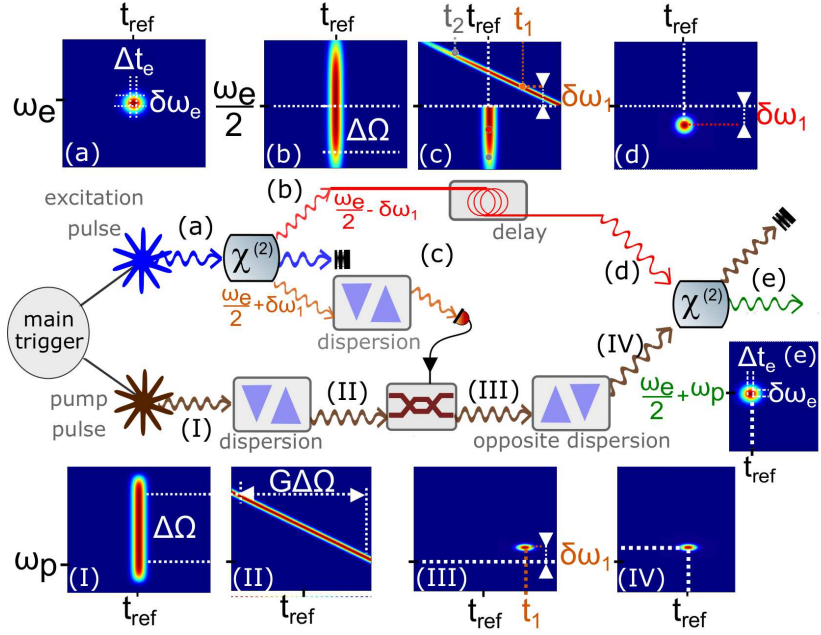}
\end{center}
\caption{Proposed scheme for the on-demand generation of single photons. Insets represent the spectral-temporal profile of the excitation pulse (a), of the bi-photon wavefunction (b,c,d),  of the output photon (e) and of the pump pulse (I-IV) at the corresponding locations.  In an illustrative example event, two pairs of photons are created. The one created at a frequency shift of $\delta \omega_1$ is detected via the detection of its Stokes photon at a timing $t_1$. A macroscopic pump pulse at a correlated wavelength is created via the synchronous carving into a broadband chirped pulse. The other pair remains undetected. Sum frequency generation between the heralded photon at $ \omega_e/2 -\delta\omega_1$ and the optical pulse at $\omega_p + \delta\omega_1$ results in a single photon at the frequency $\omega_p+\omega_e/2$ whose spectral-temporal profile is given by (e). 
}

\label{fig:2}
\end{figure}

Here we present an architecture in which the number of lasers and single photon detectors  does not scale with $N$. In fact the active resources can be reduced to one laser and one single photon detector. 
%
The principle of the new architecture is depicted in figure~\ref{fig:2}. A starting event at $t_{ref}$ triggers both an excitation pulse that will subsequently generate photon pairs and a pump pulse that will subsequently be used for frequency conversion of the single photon. The excitation pulse of frequency $\omega_e$ and duration $\Delta t_e/\sqrt{2}$ (the factor $1/\sqrt{2}$ is inserted for future convenience) generates a frequency correlated photon pair through Spontaneous parametric Down Conversion (SpDC). The two photons are separated. The heralding (or idler) photon passes through a dispersive element and is then detected by a single photon detector. The detection time $t_1$ of the heralding photon is related to its frequency $\omega_e/2 + \delta \omega_1$ through the Group Velocity Dispersion (GVD) of the dispersive element, and therefore also to the frequency $\omega_e/2 - \delta \omega_1$ of the heralded (or signal) photon. Simultaneously, the pump pulse of center frequency $\omega_p$ passes through a dispersive element with identical GVD. The signal generated at time $t_1$ is used to carve a short piece of the dispersed pump pulse at a time equal to $t_1+t_\textrm{delay}$. By adjusting the delay $t_\textrm{delay}$ one can ensure that the carved pump pulse has frequency $\omega_p + \delta \omega_1$. The carved pump pulse then passes through a dispersive element with opposite GVD so that the time of arrival of the carved pump pulse is now independent of its frequency. Sum frequency generation between the heralded photon and the carved pump pulse results in a single photon at frequency $\omega_p + \omega_e/2$.
The reader familiar with the subject will understand that there is a parallel between our method for selecting a pump pulse of frequency  $\omega_p + \delta \omega_1$ and the chirped pulse amplification~\cite{strickland1985compression} (Nobel prize 2018). Note that the excitation pulse and the pump pulse could be produced by the same pulsed laser. Note also that no fast logic is required, as the time of detection $t_1$ is directly used, through the use of dispersive elements with identical GVD, to select a pump pulse with the correct frequency. The detector deadtime ensures that another photon arriving at time $t_2$ later than $t_1$ does not trigger another frequency conversion while its partner is easily filtered out at the output.

%
The purity of active multiplexed sources, i.e. the ability to emit on demand one and only one photon, has been studied in \cite{bonneau2015effect}. Under the  assumptions of \cite{bonneau2015effect} (whose notation we copy), the purity is determined by 4 parameters: the number $N$ of modes in which photon pairs are produced, the degree of squeezing $\vert \xi \vert^2$ (assumed constant for all modes), the global collection efficiency on the idler arm $\eta_i$ which accounts for source and propagation losses as well as for the non unit efficiency of the  heralding detector, and  the overall  transmission efficiency  on the signal arm $\eta_s$ which accounts for the source brightness and propagation losses. We assume in what follows a threshold single photon detector, i.e. a detector that only differentiates between vacuum and one or more photons. 

The probability per clock-cycle that at least one idler photon is detected (the second term in Eq. \ref{Eq:p1}) is given by Eq. (12) in  \cite{bonneau2015effect}:
\begin{equation}
p_{\textrm{trig}}^{\textrm{MUX}} ( \vert \xi \vert^2, \eta_i, N) = 1 - \left( \frac{1 - \vert \xi \vert^2}{1 - (1- \eta_i)\vert \xi \vert^2} \right)^N 
\approx 1- e^{-N \eta_i \vert \xi \vert^2} ,
\label{ptrigMUX}
\end{equation}
where in the second line we have given an approximate expression valid for large $N$ and small $\eta_i$. The probability that the signal contains a single photon conditional on an idler photon being detected (the first factor in Eq. \ref{Eq:p1} - denoted  $p_{\textrm{single}}$  ) is given by Eq. (5) in  \cite{bonneau2015effect}). 
The expression for $p_{\textrm{single}}$ is complicated, but we note that for large $\eta_s$ it has the following properties:
it is maximum and equal to $\eta_s$ when $\vert \xi \vert^2=0$,  decreases with $\vert \xi \vert^2$, and equals $0$ when  $\vert \xi \vert^2=1$.

The purity, i.e. the probability per clock cycle that a heralded single photon is emitted, is given by Eq. \ref{Eq:p1}  (Eq. (13) in  \cite{bonneau2015effect}). From this expression and the properties of $p_{\textrm{single}}$ and $p_{\textrm{trig}}^{\textrm{MUX}}$ 
it follows that in order to have large purity one needs $\eta_s$ as close to $1$ as possible (low losses in the signal arm), low value of $\vert \xi \vert^2$ (i.e. low probability of producing a pair in a single mode), and large value of $p_{\textrm{trig}}^{\textrm{MUX}}$. To achieve the latter, one can tolerate large losses in the idler arm and small $\vert \xi \vert^2$, provided these are compensated by a sufficiently large value of $N$. An essential feature of the present proposal, already emphasized above, is that the losses $(1-\eta_s)$ in the signal channel do not scale with $N$ (and can be kept very low). The number $N$ of frequency bins is upper bounded by the GVD of the dispersive element (denoted $G$) times the bandwidth $\Delta \Omega$ (in rad/s, i.e. angular frequency) of the signal and idler photons divided by the time jitter $\delta t_d$ of the single photon detector:
\begin{equation}
N \leq \frac{ G  \Delta \Omega }{ \delta t_d} \ .
\label{EqN}
\end{equation}
It is also upper bounded by the bandwidth of the excitation pulse
\begin{equation}
N \simeq \Delta \Omega \frac{\Delta t_e}{\sqrt{2}}\ .  
\label{NDeltate}
\end{equation}
The bandwidth $\Delta \Omega$ is either the bandwidth of the SpDC or the acceptance bandwidth of the dispersive element, whichever is smaller. Given Eq. \ref{ptrigMUX} and Eq. \ref{EqN}, it follows that the product $G$ times $\Delta \Omega$ times the collection efficiency (1 minus losses) of the dispersive element is a critical parameter that must be as large as possible. In this regard a Dispersion Compensation (DC) module (based on chirped fiber Bragg gratings) seem to be the best choice. In what follows we  take the parameters of a commercially available DC module\cite{proximon} : $G=8000$ ps$^2$, Bandwidth$=\Delta \Omega=2\pi*1$~THz, Losses$=3$~dB.  As this bandwidth is  less than the bandwidth that can be reached with SpDC (which can exceed  $2\pi*10$~THz \cite{Vanselow:19,Kaneda:16,ngah2015ultra}), we can suppose that the full bandwidth  is used and that the degree of squeezing $\vert \xi\vert^2$ is constant over the bandwidth.

The time jitter $\delta t_d$ of single photon detectors ranges from as low as $10$ ps for commercially available superconducting detector to $100$s picoseconds for single photon silicon avalanche photodiodes, while efficiency ranges from 20 to 90 \%.  Assuming $\delta t_d = 10$ ps, we find from Eq. \ref{EqN} that $N\leq 5000$. It is this large value, easily accessible with current technology, which makes active heralding based on frequency mutliplexing so appealing.

The sum frequency generation is realised between the signal photon and the carved pump pulse. 
This can be realised via sum frequency generation~\cite{kumar1990quantum} in a $\chi^{(2)}$ material or via Bragg scattering four wave mixing in a $\chi^{(3)}$ material. In both cases,  frequency conversion that is both noiseless and has near unit probability of success has been reported \cite{clemmen_ramsey_2016,albota2004efficient}. 
Table \ref{tab:freqconv} presents the efficiency and bandwidth reported in some recent experiments. These do not yet match the requirements of the present proposal, but the field is evolving at a fast pace, in particular with the development of nanophotonic $\chi^{(2)}$ waveguides~\cite{jankowski2020ultrabroadband,wang2020efficient,li2016efficient}.

We now  consider the wavefunction of the signal photon, and the amplitude of the carved pump pulse.
After SpDC,  the wavefunction of the signal and idler photons, assuming that a single pair is produced, is (up to normalisation which we omit in all expressions below)
\begin{equation}
\vert \psi \rangle_{si} =  \int_{\Delta \Omega} d\omega_d \int d\omega_s \ 
e^{-\left( \omega_s - \omega_e\right)^2\frac{\Delta t_e^2}{2} }
\vert \frac{\omega_s}{2}+\omega_d\rangle_i  \vert \frac{\omega_s}{2}-\omega_d\rangle_s  \ ,
\end{equation}
where we have assumed that the excitation pulse has a gaussian shape.
When the idler photon passes through the dispersive element its wavefunction is multiplied by $e^{i(\frac{\omega_s}{2}+\omega_d-\omega_e/2)^2\frac{G}{2}}$. Hence, after the detection of the idler photon at time $t_1$, the wavefunction of the signal photon is
\begin{equation}
\psi_s(t)=e^{-\frac{t^2}{2 \Delta t_e^2}}
e^{-i\frac{(t-t_1)^2}{2G} } e^{-i \frac{ \omega_e (t+t_1)}{2}}
\end{equation}
where $t$ is related to the position along the propagation direction by $t=z/v$ with $v$ the propagation velocity. The signal photon wavefunction thus has a wavepacket of duration $\Delta t_s = \Delta t_e/\sqrt{2}$, frequency $\omega_e/2 - t_1/G$, and chirp  $1/G$. (Note that here and below by duration of a wavefunction or classical pulse we mean the standard deviation of $t$, and by spectral width the standard deviation of $\omega$. As a consequence we have $\Delta t \Delta \omega \geq 1/2$ with the inequality saturated for unchirped gaussian wavepackets).
 
The signal generated at time $t_1$ is used to carve into a gaussian shape (with duration $\tau$) a short piece of the dispersed pump pulse. The complex amplitude of the pump pulse, after carving  but before passing through the second dispersive element, is
 \begin{equation}
A_{pump}(t) =A e^{-i  \omega_p t} e^{-i\frac{t^2}{2G} }  e^{-\frac{(t-t_1)^2}{2\tau^2} } \ ,
 \end{equation}
and after passing through the second dispersive element is
 \begin{equation}
A'_{pump}(t)=A e^{-i  (\omega_p + t_1/G) t} e^{-\frac{t^2\tau^2}{2G^2} }  e^{i\frac{t^2}{2G} } \ ,
 \end{equation}
The pump pulse $A'_{pump}(t)$ therefore has duration $\Delta t_{pump}= G/\sqrt{2}\tau$, spectral width 
$\Delta \omega_{pump} = \sqrt{\frac{1}{2}(\frac{\tau^2}{G^2} + \frac{1}{\tau^2})}$, and is centered on the frequency $\omega_p+ t_1/G$. The two terms in  $\Delta \omega_{pump} $ come from two contributions: 1) carving the pump pulse selects a piece of the pump with spectral width equal to $\tau/\sqrt{2}G$; and 2) carving a pulse of duration $\tau$ implies an intrinsic spectral width given by $(\tau \sqrt{2})^{-1} $.

The wavefunction of the converted single photon $\psi_h$ can be estimated as the product of the signal  wavefunction and the pump amplitude:
%
\begin{equation}
\psi_h(t) \simeq  \psi_s(t) A'_{pump}(t)=e^{-(\frac{1}{ \Delta t_e^2} +\frac{\tau^2}{G^2}  )\frac{t^2}{2}}
 e^{-i( \frac{ \omega_e }{2} + \omega_p)t} e^{-i \varphi(t_1)} \ . 
\end{equation}
We see that,  except for the unimportant phase $\varphi(t_1)$, the wavefunction of the heralded photon is independent of $t_1$ and its center frequency is $\frac{ \omega_e }{2} + \omega_p$.
Note that   the chirps have canceled and the heralded photon is Fourier transform limited
with duration $\Delta t_h$ and spectral width $\Delta \omega_h$ given by 
\begin{equation}
\Delta \omega_h = 1/(2 \Delta t_h ) = \frac{1}{2} \sqrt{\frac{1}{ \Delta t_e^2} +\frac{\tau^2}{G^2} } \ . 
\label{Eq:Deltaomegah}
\end{equation}

The present proposal thus requires frequency conversion of a single photon using a pulsed pump. Work on frequency conversion has so far focused on quasi-CW pumps. Indeed a pulsed pump can result in lower conversion efficiency when its instantaneous intensity varies on the scale of the single photon's coherence time~\cite{brecht2015photon}. In order to circumvent this difficulty, and  to be as close as possible to the case of a cw pump, one needs to choose the duration of the pump pulse much longer than the duration of the signal wavefunction: 
$\Delta t_{pump}= G/\sqrt{2}\tau \gg \Delta t_e/\sqrt{2}$, hence
\begin{equation}
\Delta t_e \ll \frac{G}{\tau} \ .
\label{Eq:condteGtau}
\end{equation} 

We now consider the indistinguishability of the heralded photon which is affected by the
 jitter $\delta t_d$ of the single photon detector. Indeed this jitter results in an uncertainty on the frequency of the pump pulse which we can write as $\omega_p + \delta \omega_1 \pm \delta t_d / G$. Therefore the frequency of the heralded photon, after sum frequency generation, will be 
$\omega_p + \omega_e/2 \pm \delta t_d / G$.

This classical noise on the frequency of the heralded photon should be compared to the intrinsic spectral width of the heralded photon. Indeed, the effect of $\delta t_d$ on the indistinguishability is determined by the spectral width $\Delta \omega_h$ of the heralded photon.  
We can estimate the effect of the jitter by writing the heralded photon wavefunction as
\begin{equation}
\psi_h(t; t_d)\simeq e^{-\frac{t^2}{ 2 \Delta t_h^2} }
 e^{-i( \frac{ \omega_e }{2} + \omega_p + \frac{t_d}{G})t} \ ,
\end{equation}
where $t_d$ is the jitter induced by the detector.
The corresponding quantum state is denoted $\vert \psi_h (t_d)\rangle$ and
the density matrix of the heralded photons is 
$\rho_h = \int dt_d \ P(t_d) \vert \psi_h(t_d)\rangle \langle  \psi_h(t_d)\vert $
where $P(t_d)$ is a gaussian of width $\delta t_d$. The visibility of the dip in a HOM experiment is computed to be
\begin{equation}
V = \textrm{Tr} \left( \rho_h^2 \right) = \left( 1 + \frac{2\delta t_d^2 \Delta t_h^2}{G^2} \right)^{-\frac{1}{2}} \ . 
\label{Eq:V}
\end{equation}

As example of experimental parameters, we consider the DC module described above. We suppose that the jitter $\delta t_d = 10$~ps and that the carved pump pulse has duration $\tau = 10$~ps.
(The pump pulse itself is much shorter, so that its spectrum covers the whole bandwidth $\Delta \Omega$ of the DC module). The duration of the pump pulse after carving and recompression is then $\Delta t_{pump}= \frac{G}{\sqrt{2}\tau}= 570$~ps.
We take an excitation pulse which is $10$ times shorter, so that  parameter $\Delta t_e=80$~ps, which ensures that inequality Eq. \ref{Eq:condteGtau} is satisfied. The visibility Eq. \ref{Eq:V} will then be higher than $99.8\%$.
Using a short excitation pulse will result in a number of usable frequency bins approximately given by Eq. \ref{NDeltate}, which yields $N = 500$. 
Assuming collection efficiency on the signal arm $\eta_s=0.85$ and on the idler arm $\eta_i=0.3$, we find from Eq. \ref{Eq:p1} that the optimal squeezing parameter is $\vert \xi \vert^2 = 0.031$  yielding a probability per clock-cycle that a single photon is produced $p_{1} = 0.81$.

In summary, we have proposed a method to realise active multiplexing of heralded single photons based on frequency multiplexing. The proposal leverages the  high dimensionality of frequency entangled  photons pairs. It is economical in terms of experimental resources, requiring two (or even a single) pulsed laser. The key parameter that determines the quality of the source is the collection efficiency on the signal arm $\eta_s$. Making  $\eta_s$ as large as possible will require further work on frequency conversion of single photons, particularly in the regime of a pulsed and chirped pump used here.

\begin{table}
\begin{center}
\begin{tabular}{|c|c|c|c|}  
    \hline
    Nonlinear systems  & $\eta$ & BW  \\
    \hline
    SFG in PPLN crystal \cite{albota2004efficient} & 93 \%  &  20 GHz   \\
                                                         \hline 
    SFG in PPLN planar-wg \cite{li2017broadband} & 0.1\% &  $0.6$ THz  \\  \hline   
    FWM in  optical fiber \cite{clemmen2018all} & $>80 \% $  & 1.2 THz   \\ \hline
\end{tabular}
\caption{\label{tab:freqconv} Characteristic of some frequency conversion experiments. SFG: sum-frequency generation, FWM: four-wave mixing,  $\eta$: experimentally reported ratio of up-converted  to incident photons fluxes, BW: (-1~dB)-conversion bandwidth specified for a fixed pump wavelength in the case of SFG and with the second pump tunable in the case of FWM.
}
\end{center}
\end{table}  
\section*{Funding.} 
Stéphane Clemmen is a research associate of the Fonds de la Recherche Scientifique – FNRS.

\bibliographystyle{unsrt}
\bibliography{references}

\end{document}